\documentclass[12pt,a4paper]{article}

\usepackage[font=scriptsize]{caption}
\usepackage{titlesec}
\usepackage{multicol}
\usepackage{flushend}
\usepackage{amsfonts}
\usepackage{setspace}
\usepackage{balance}
\usepackage{nomencl}
\usepackage[numbers]{natbib}
\usepackage{amsmath}
\usepackage{amsthm}
\usepackage{amsmath}
\usepackage{amssymb}
\usepackage{float}
\usepackage{stmaryrd}
\usepackage{stackrel}
\usepackage{url}
\usepackage{color,hyperref}
\usepackage{subfig}
\usepackage{graphicx}
\usepackage{geometry}
\begin{document}



\title{Geodesics Structure and Thermodynamic Properties of Gaussian Black Hole in Quadratic Ricci Scaler Gravity}

\maketitle


\begin{center}
\author{{M. Haditale} and {B. Malekolkalami}}

\thanks{Faculty of Science, University of Kurdistan, Sanandaj, P. O. Box 416, Iran (email: maryam.haditale@uok.ac.ir and
b.malakolkalami@uok.ac.ir)}
\end{center}

\begin{abstract}
The geodesic structure and thermal properties of  Gaussian Black Holes (\textbf{GBH})s in modified  and Einstein gravities are studied and compared. In the geodesic part, motion of a test particle (massive and massless) are discussed, specially properties of the circular motion are considered. In the  thermodynamic part,  the  mass,  entropy and  temperature functions  are considered and discussed. The local and global stability is also analyzed through the  Heat Capacity (\textbf{HC}) and Gibbs Energy (\textbf{GE}). The results show the thermodynamic differences are more than geodesic ones in the two theories of gravity with the note that the  modified gravity is more consistent with the physical world.
\end{abstract}


\section{Introduction}\label{sec-1}
The theory of General Relativity, developed by Einstein, effectively describes gravitational interactions and has been validated through numerous experiments \cite{Airy1}. In 1998, it was discovered that the universe is expanding at an accelerating rate, a finding supported by observational evidence from type Ia supernovae \cite{Airy2}, measurements of the Cosmic Microwave Background (\textbf{CMB}) \cite{Airy3}--\cite{Airy5}, and studies of the large-scale structure \cite{Airy6}.
In General Relativity, the acceleration of the universe is attributed to Dark Energy (\textbf{DE}), which is supported by both theoretical and observational evidence. However, two key issues persist: the difficulty in deriving vacuum energy from quantum field theory and the mismatch between the quantities of DE and Dark Matter (\textbf{DM}) in the universe \cite{Airy7}. As a result, researchers are investigating modifications to gravity on a cosmological scale as a potential alternative to the DE model \cite{Airy8}.
Research into Black Holes (\textbf{BH}s) in anti-de Sitter (\textbf{AdS}) and de Sitter (\textbf{dS}) spacetimes is exploring modified gravity theories, such as $f(R)$, to address issues in classical General Relativity \cite{Airy9}. Einstein-based gravity theories, particularly those involving higher curvature corrections like $R + R^{n}$, are gaining attention. In string theory, these corrections naturally arise and while including quadratic curvature terms ($n = 2$) can make the theory renormalizable, it may also introduce problematic ghost-like states from certain curvature tensor squares \cite{Airy10}.
The $R + R^{2}$ theory of gravity is a ghost-free alternative to Einstein's general relativity, incorporating an additional scalar degree of freedom, $\phi$. It has garnered interest in cosmology for its ability to model a smooth transition from a decelerated phase to a flat Minkowski phase, providing insights into early inflationary scenarios  \cite{Airy11}, \cite{Airy12}. Recent studies have also examined the $R + R^{2}$ inflation scenario in the context of supergravity, enhancing our understanding of gravitational dynamics \cite{Airy13}, \cite{Airy14}.\newline
The term $R^{2}$ gravity is frequently referenced in research on higher curvature gravity and its solutions within linear Einstein theory, as well as in supergravity and string theory \cite{Airy15}. It is characterized by two main features: it is the only theory with quadratic terms in the curvature tensor and it provides a gravitational framework with a fixed interaction structure, unlike standard low-energy models that exhibit scale variability. The $R^{2}$ gravity in Einstein's framework, describes an action with a cosmological constant, leading to dS or AdS spaces due to global scale symmetry. Additionally, static spherically symmetric solutions in scale-invariant $R^{2}$ gravity have been found, showing alignment with solutions from Einstein's theory. Overall, $R^{2}$ gravity presents a variety of vacuum static spherically symmetric solutions \cite{Airy16}.\newline
This theory possesses two notable features: first, it uniquely avoids the presence of a regular quadratic ghost in the curvature tensor. Second, it is a scale-invariant gravitational theory with a complex interaction structure, akin to the scale invariance observed in the low-energy (supersymmetric) Standard Model, which lacks a Higgs mass term at the classical level.

Many physicists believe that the emergence of singularities in mathematical physics is due to mathematical calculations because they cannot exist in the real world. For this reason, free singularity theories as regular BHs  has attracted the attention of many researchers in recent years. Therefore, the theory of regular  BHs has been welcomed by  many of people. An example of BHs with regular geometry is GBH. In this article we want to examine some geometric and physical properties of their corresponding spacetimes, in framework of $R^{2}$ gravity.

In 1975, Stephen Hawking revealed that BHs can evaporate and emit thermal radiation similar to black body radiation \cite{Airy17}, a process linked to quantum fluctuations near the event horizon. This phenomenon, where particle-antiparticle pairs are created, contributes to the field of Quantum Gravity, which merges Quantum Mechanics and General Relativity. Understanding the statistical mechanics of BHs is crucial for advancing theories of quantum gravity, such as String Theory and Loop Quantum Gravity, despite the challenges in experimental validation \cite{Airy18}-\cite{Airy21}. Effective theories, particularly Quantum Field Theory in Curved Space and Noncommutative Geometry, are essential for analyzing BH evaporation and exploring quantum BHs \cite{Airy22}-\cite{Airy24}. Noncommutative Geometry enhances the renormalization of theories at short distances and provides a framework applicable to both String Theory and Loop Quantum Gravity, offering new insights into the nature of spacetime and physical phenomena \cite{Airy25}-\cite{Airy27}. In mathematical physics, noncommutative field theory is an application of noncommutative mathematics to the spacetime  in which the coordinate functions ${X}^{\mu }$ are noncommutative. One commonly studied version of such theories can be presented by:
\begin{center}
${[{{X}^{\mu }},{{X}^{\nu }}]=i{{\alpha }^{\mu \nu }}}$,
\end{center}
where ${{\alpha }^{\mu \nu}}$ is an antisymmetric matrix with dimensions of length squared. In the realm of Noncommutative Geometry, considering Einstein gravity, the momentum--energy tensor yields a Gaussian distribution of matter. To be more precise, it is shown that in an asymptotically flat  spacetime, the noncommutativity effects can remove the point--like structures replaced by smeared objects described by a Gaussian matter distribution \cite{Airy28}.\\
Our objective in this research is to compare the properties of the Gaussian Black Hole (\textbf{GBH}) solution in $R^{2}$ gravity with that in Einstein's theory. We looking for aymptotically flat solutions  for GBH spacetime and analyze it's geodesic structures around it, as well as their thermodynamic properties, including entropy, mass and temperature. Also, we evaluate the local and global stability through the heat capacity (HC) and Gibbs energy (GE) criteria, specially.\\
The structure of the paper is as follows: in Sect. \ref{sec-2}, we present GBH solution in $R^{2}$ gravity and briefly introduced in Einstein's gravity. In Sect.\ref{sec-3}, the geodesic equations for photon and massive particle are discussed  and shown by corresponding plotted paths. In Sect.\ref{sec-4}, the thermodynamic  functions as entropy, temperature ... are investigated. The main conclusions are given in Sect.\ref{sec-5}.
\section{GBH in $F(R)$ gravity}\label{sec-2}
\begin{figure}[H]
\centering
\includegraphics[width=12cm]{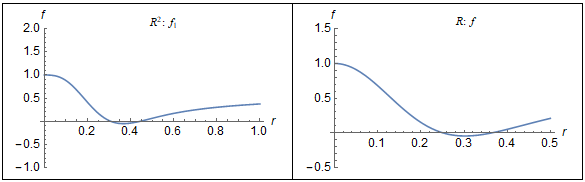}
\caption{The metric coefficients (\ref{eq10}) and (\ref{eq11}) for $M=50\sqrt{\alpha}$ and $\alpha=0.01$  are shown in the left and right panels, respectively.}
\label{fig1-1}
\end{figure}

The modified gravity $F(R)$  action, when incorporating a matter field characterized by the Lagrangian density, can be described in the following  comprehensive form \cite{Airy29}:
\begin{equation}
S=\int{\Big(\frac{1}{16\mathop{\mu }^{2}}F(R)-\mathop{L_{m}}\Big)}\mathop{\sqrt{-g}d^{4}x},
\label{eq1}
\end{equation}
where  $L_{m}$ is   the matter Lagrangian considered here as  Gaussian distribution whose  corresponding   Energy--Momentum tensor  is given by the following non--zero components \cite{Airy28}:
\begin{align}
\mathop{T_{0}^{0}}=\mathop{T_{r}^{r}}=-\rho (r) &\nonumber\\
\mathop{T_{\theta }^{\theta }}=\mathop{T_{\phi }^{\phi }}=-\rho (r)-\frac{r}{2}\frac{\partial \rho (r)}{\partial r},
\label{eq2}
\end{align}
where
\begin{equation}
\rho (r)=\frac{M}{\mathop{(4\pi \alpha )}^{\frac{3}{2}}}{e}^{\frac{-{r}^{2}}{4\alpha }}.
\label{eq3}
\end{equation}
Varying the action (\ref{eq1}) with respect to the metric, we get the generalized gravitational field equations as follows:
\begin{equation}
F'(R)R_{\mu }^{\beta }-\frac{1}{2}F(R)g_{\mu }^{\beta }-\Big({{\nabla }_{\mu }}{{\nabla }^{\beta }}-g_{\mu }^{\beta }{{\square }^{2}}\Big)F'(R) =T_{\mu }^{\beta },
\label{eq4}
\end{equation}
where the prime stands for derivative with respect to $R$. By replacing $F(R)=R^{2}$ in (\ref{eq4}), we obtain:
\begin{equation}
2RR_{\mu }^{\beta }-\frac{1}{2}{{R}^{2}}-2\Big({{\nabla }_{\mu }}{{\nabla }^{\beta }}-{{\square }^{2}}\Big)R =T_{\mu }^{\beta }.
\label{eq5}
\end{equation}
The metric  solution for the field equations (\ref{eq5}) is considered as   spherically symmetric solutions of the form below
\begin{equation}
\mathop{ds^{2}}=-f(r)\mathop{dt^{2}}+\frac{\mathop{dr^{2}}}{f(r)}+\mathop{r^{2}}\mathop{d\Omega ^{2}},
\label{eq6}
\end{equation}
where the  metric  function $f(r)$ must satisfy the asymptotically flat condition. Given the energy--momentum tensor (\ref{eq2}) and substituting the  metric (\ref{eq6})  into the field equations (\ref{eq5}) and solving them, we get the following differential equation:
\begin{align}
f''^{2}&+\frac{8f''f'}{r}+\frac{8f'^{2}}{r^{2}}+\frac{2ff''}{r^{2}}+\frac{4ff'}{r^{3}}-\frac{2f''}{r^{2}}-\frac{4f'}{r^{3}}-\frac{r^{2}f''^{2}}{2}-4rf'f''\nonumber\\&
-2ff''+2f''-\frac{8ff'}{r}+\frac{8f'}{r}+\frac{4f}{r^{2}}-8f'^{2}-\frac{2}{r^{2}}=-\frac{M}{\mathop{(4\pi \alpha )}^{\frac{3}{2}}}{e}^{\frac{-{r}^{2}}{4\alpha }},
\label{eq7}
\end{align}
where the prime stands for derivative with respect to $r$. Given the complexity of this differential equation, it cannot be solved using manual methods. The following solutions can be obtained using software tools:
\begin{align}
{{f}_{1-R^{2}}}(r)=&1+\frac{c_{1}}{r}+\frac{c_{2}}{{{r}^{2}}}+\frac{10M\alpha }{{{(4\pi \alpha )}^{\frac{3}{2}}}}{{e}^{-\frac{{{r}^{2}}}{4\alpha }}}r+\frac{40M{{\alpha }^{2}}}{{{(4\pi \alpha )}^{\frac{3}{2}}}}\frac{{{e}^{-\frac{{{r}^{2}}}{4\alpha }}}}{r}\nonumber\\&-\frac{M}{4\pi }\Big(1+\frac{12\alpha }{{{r}^{2}}}\Big)Erf\Big[\frac{r}{\sqrt{4\alpha }}\Big],
\label{eq8}
\end{align}
and
\begin{align}
{{f}_{2-R^{2}}}(r)=&1+\frac{c_{3}}{r}+c_{4}{{r}^{2}}+\frac{6M\alpha r}{{{(4\pi \alpha)}^{\frac{3}{2}}}}{{e}^{-\frac{{{r}^{2}}}{4\alpha }}}r+\frac{16M{{\alpha }^{2}}}{{{(4\pi \alpha )}^{\frac{3}{2}}}}\frac{{{e}^{-\frac{{{r}^{2}}}{4\alpha }}}}{r}\nonumber\\
&-\frac{M}{8\pi \alpha }\Big(2\alpha +\frac{{{r}^{2}}}{3}\Big)Erf\Big[\frac{r}{\sqrt{4\alpha }}\Big],
\label{eq9}
\end{align}
where $c_i$ are integration constants and the error function is given by:
\begin{equation}
Erf(x)={\frac {2}{\sqrt {\pi }}}\int _{0}^{x}e^{-t^{2}} {d}t. \nonumber
\end{equation}
The metric solution (\ref{eq8}) satisfies the asymptotically flat condition and  metric (\ref{eq9}) not, then we proceed only with the  solution (\ref{eq8}).\\This  solution appears to have a singularity at origin, however it will be regularized, by choosing the   integration constants ($c_{1}$, $c_{2}$) as\footnote{It can be easily verified.}:
\begin{equation}
c_1= -16 \alpha^{2}\rho_0= -\frac{2M}{\pi}\sqrt{\frac{\alpha}{\pi}},\hspace{1.5cm} c_2=0, \nonumber
\end{equation}
where $\rho_0 = \frac{M}{\mathop{(4\pi \alpha )}^{\frac{3}{2}}}$. By replacing the above constants  in (\ref{eq8}),  we have:
\begin{align}
f_{R^{2}}(r)=&1-\frac{2M}{\pi}\sqrt{\frac{\alpha}{\pi}}\Big(\frac{1}{r}\Big)+\frac{10M\alpha }{{{(4\pi \alpha )}^{\frac{3}{2}}}}{{e}^{-\frac{{{r}^{2}}}{4\alpha }}}r+\frac{40M{{\alpha }^{2}}}{{{(4\pi \alpha )}^{\frac{3}{2}}}}\frac{{{e}^{-\frac{{{r}^{2}}}{4\alpha }}}}{r}\nonumber\\&-\frac{M}{4\pi }\Big(1+\frac{12\alpha }{{{r}^{2}}}\Big)Erf\Big[\frac{r}{\sqrt{4\alpha }}\Big],
\label{eq10}
\end{align}
where for simplicity the subscript 1 is omitted. \vspace{3mm}\\
\textbf{GBH in Einstein's Gravity:}\\
For further comparisons, it is essential to present the metric associated with the Gaussian distribution within the Einstein gravity which it  is shown that  the  metric function takes the following form \cite{Airy28}:
\begin{equation}
f_{R}(r)=1+\frac{c}{r}+\frac{M \alpha}{{(4\pi \alpha )}^{\frac{3}{2}}}{e}^{-\frac{{{r}^{2}}}{4\alpha }}-\frac{M}{4\pi r }Erf\Big[\frac{r}{\sqrt{4\alpha}}\Big],
\label{eq11}
\end{equation}
where for regularity, the integration constant must be zero, that is  $c=0$. The metric functions (\ref{eq10}) and (\ref{eq11}) are plotted versus $r$ in Fig.\ref{fig1-1},  they are qualitatively similar.
\section{The Geodesics}\label{sec-3}
In this section, we examine the geodesics paths  around the GBH spacetime. One of the useful concepts in this regard is the effective potential energy, so let us give a brief statement of this concept.

\begin{figure}[]
  \centering
  \includegraphics[width=9cm]{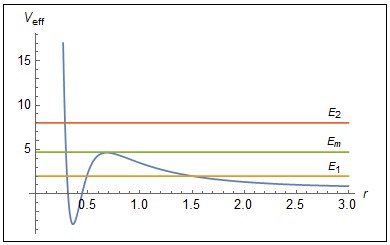}
  \caption{The effective potential graph versus $r$ for massive particle with values of $\alpha=0.01$ and $M=50\sqrt{\alpha}$. The horizontal lines represent the energy levels.}
  \label{fig2-2}
\end{figure}

In General Relativity, the effective potential is a helpful concept for analyzing the motion of particles around a central mass. It combines gravitational and centrifugal forces into a single potential-energy-like function, simplifying orbit analysis by reducing the system to one dimension. The total energy determines its shape, a characteristic tool  in  orbit's classification, especially to determine the stability of circular orbits. By plotting the effective potential, one can determine the allowed regions of motion for a particle and identify circular orbits. Also, it is a valuable tool for analyzing the motion of a free particle in the equatorial plane of a spherically symmetric attraction.\\
For spherically symmetric spacetime (\ref{eq6}), the effective potential is given by \cite{Airy30}:\\
\begin{equation}
{{V}_{eff}}=\Big(\epsilon+\frac{{{L}^{2}}}{{{r}^{2}}}\Big)f(r),
\label{eq12}
\end{equation}
where $L$ is generalized momentum corresponding to the $\phi$ coordinate and $\epsilon = 0$ and $\epsilon = 1$ correspond to the cases of  photon and  massive particle, respectively.
Given a metric function (\ref{eq10}), we can plot the effective potential (\ref{eq12}) versus radial coordinate $r$, as shown in Figure \ref{fig2-2}, for the values:  $M=50\sqrt{\alpha}$, $\alpha=0.01$ and $\epsilon=1$ (massive particle)\rlap.\footnote{It should be noted, the graphs of the effective potential (12) for the  massless particle and metric function (11) are qualitatively similar to Figure 2.}

As the figure shows there are two extremum points. First is a minimum   corresponding to a stable circular orbit between the inner and outer horizons and second is an maximum  corresponding to an unstable circular orbit outside the outer horizon.  Finally, we note that outside the outer horizon for energies smaller than $E_m$,  there is always a turning point, meaning that the particle can escape to infinity.
We now return to the determination of geodesics.\\
A key point to astrophysical application of a gravitational metric is how point particles and photons move in it.
In  theory of relativity, gravitation is not a force but a property of spacetime geometry where a test particle and light move in response to the geometry of the spacetime.
In fact, one of the ways to exploring the curved spacetimes is  reviewing the nature of the motion of freely falling particles\footnote{Freely falling particles are those particles that are free from any effects except curvature of spacetime.} in them. The paths of freely falling particles are the same as geodesics and our attempt in this section is to provide a picture of geodesic structure of  GBH spacetime. The full geodesic equation is
\begin{equation}
\mathop{{\ddot{x}}^{\alpha }}+\mathop{\Gamma _{\beta \nu }^{\alpha }}\mathop{{\dot{x}}^{\beta }}\mathop{{\dot{x}}^{\nu }}=0,
\label{eq13}
\end{equation}
where $\Gamma _{\beta \nu }^{\alpha}$ are affine connections and dots denote derivative with respect to a scalar parameter\footnote{Although not necessarily, but for material particles, proper time is usually considered as a parameter.}. The spherical static  symmetry  geometry allows to study  the case of a particle moving in the equatorial plane  $( \theta=\pi/2)$ without loss of generality. By imposing this, the geodesic equations (\ref{eq13}) reduce to:
\begin{equation}
\ddot{t}+\left(\frac{f'(r)}{f(r)}\right)\dot{t}\dot{r}=0,
\label{eq14}
\end{equation}
\begin{equation}
\ddot{r}+\frac{1}{2}f(r)f'(r)\mathop{{\dot{t}}^{2}}-\frac{1}{2}\frac{f'(r)}{f(r)}\mathop{{\dot{r}}^{2}}-rf(r)\mathop{{\dot{\phi }}^{2}}=0,
\label{eq15}
\end{equation}
\begin{equation}
\ddot{\phi }+\frac{2}{r}\dot{r}\dot{\phi }=0.
\label{eq16}
\end{equation}
\begin{figure}[]
\centering
\includegraphics[width=12cm]{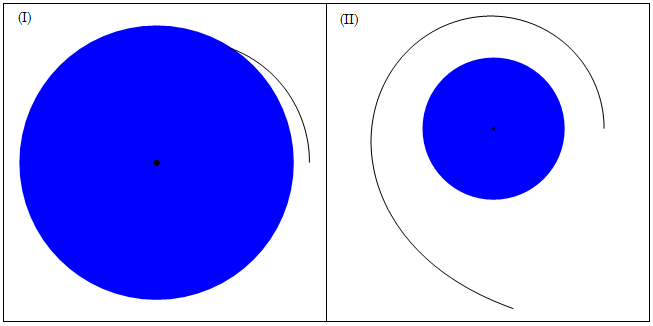}
\caption{Geodesic paths of massive and massless particles in $R^{2}$ gravity (corresponding to metric (\ref{eq10})) for $\alpha=0.01$ and $M=50\sqrt{\alpha}$.\\
I: $r_{0}=0.6$ and $r_{h}=0.45$; II: $r_{0}=1$ and $r_{h}=0.45$; The radius of the blue disk represents the event horizon $r_{h}$.}
\label{fig3-3}
\end{figure}
\begin{figure}[]
\centering
\includegraphics[width=12cm]{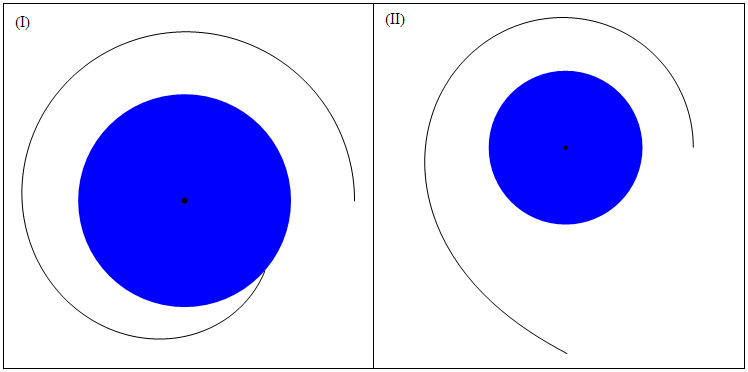}
\caption{Geodesic paths of massive and massless particles in $R$ gravity (corresponding to metric (\ref{eq11})) for $\alpha=0.01$ and $M=50\sqrt{\alpha}$.\\
I: $r_{0}=0.6$ and $r_{h}=0.36$; II: $r_{0}=0.6$ and $r_{h}=0.36$.}
\label{fig4-4}
\end{figure}
Given the nonlinearity of the  system of differential equations above, it isn't possible the exact solutions for $t, r, \phi$ as explicit functions of  scalar parameter. Fortunately, this is not necessary for our purpose (i. e.  showing the  paths of motion in the equatorial plane) because it can be done by  using numerical instructions. To utilize this method, the boundary conditions must be known, which are considered as follows:
\begin{center}$r(0)= r_0=0.6,\hspace{2mm} \phi(0)= 0, \hspace{2mm} \dot{r}(0)=0$ and $L=\mathop{r}^{2}\dot{\phi}=3.$\end{center}
The outputs of the software calculations for plotting  paths are shown in the Figures \ref{fig3-3}  and \ref{fig4-4} for the metric functions (\ref{eq10}) and (\ref{eq11}), respectively.
As the figures show, the geodesic paths are qualitative similar as expected, on the  qualitative similarity of the corresponding  potential diagrams (Fig.\ref{fig3-3}).  But,  there is a quantitative difference (rather significant) between  the left panels (in Fig. \ref{fig3-3} and Fig. \ref{fig4-4}).  They  indicate that in the modified gravity the particle takes a shorter path in falling into the horizon,  physically meaning  that, the  gravity is stronger in $R^{2}$ framework. The right panels of Fig. \ref{fig3-3} and Fig. \ref{fig4-4} show the situation where $r_{0}$ is chosen to be just less than twice the value of $r_{h}$ (the event horizon of BH).  Also, it is noted that the gravity becomes stronger  for increasing $\alpha$ parameter, that is it causes a shorter path for particles falling into the BH.
\subsection{Circular Orbits}\label{section-3}
Circular orbits are the simplest type of orbit in celestial mechanics, in which a rotating object remains at a constant radius while moving around a gravitational mass. Circular orbits are particularly important in the equatorial plane of the celestial objects. For example, they are relevant to the structure of accretion disks around astrophysical BHs.\\
As mentioned above, a circular stable (unstable) orbit corresponds to a minimum (maximum) point of effective potential profile. In connection with the current discussion, referring to the Fig. \ref{fig2-2}, it can be seen, there is a stable circular orbit inside the horizon and unstable orbit outside the horizon. In below, the radius and frequency of  circular orbits out side the horizon are calculated.

\begin{figure}[]
  \centering
  \includegraphics[width=12cm]{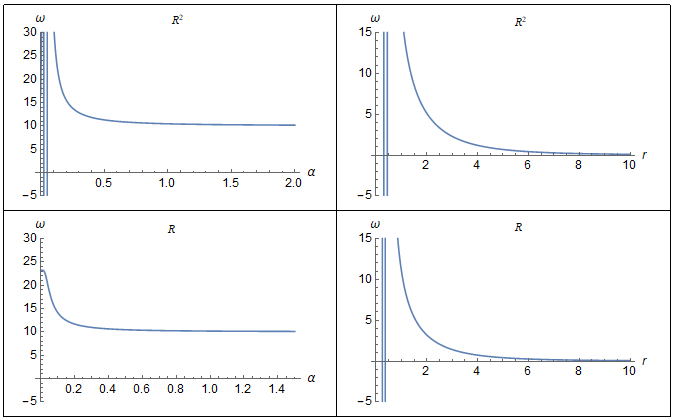}
  \caption{Orbital frequency $\omega$ versus $\alpha$ and $r$ in $R^{2}$ and $R$ gravities for $M=50\sqrt{\alpha}$, $L=3$, $E=4.9$ and $\epsilon=0,1$.}
  \label{fig5-5}
\end{figure}

For this,  we begin with the equation $dV_{eff}/dr=0$ which by  replacing from (\ref{eq12}) reads:
\begin{equation}
\frac{d{{V}_{eff}}}{dr}=-\frac{2{{L}^{2}}f(r)}{{{{{r}^{3}}}}}+\left(\frac{L^{2}}{r^{2}}+\epsilon\right){f}'(r)=0,
\label{eq17}
\end{equation}
which after simplifying becomes
\begin{equation}
\frac{{f}'(r)}{f(r)}=\frac{2{{L}^{2}}}{L^{2}r+\epsilon r^3}.
\label{eq18}
\end{equation}
Now, noting that for a circular motion ($r=cte$), one has $\dot{r}=\ddot{r}=0$, $\dot{\phi}=\omega$ and therefore by substituting these into (\ref{eq15}), we get
\begin{equation}
\frac{1}{2}f'(r)\mathop{{\dot{t}}^{2}}-r \mathop{\omega^{2}}=0,
\label{eq19}
\end{equation}
on the other hand, from equation (\ref{eq14}), it is  easily obtained  $\dot{t}=\frac{E}{f(r)}$ where by substituting it into (\ref{eq19}), we obtain
\begin{equation}
\mathop{\omega^{2}}=\frac{E^2 f'(r)}{2f^2(r)},
\label{eq20}
\end{equation}
finally by eliminating the derivative of the metric  function ($f'$) from  equations (\ref{eq18}) and (\ref{eq20}), the orbital  frequency can be obtained as function of the orbital radius and the Gaussian distribution parameter $\alpha$, that is:
\begin{equation}
\mathop{\omega^{2}}=h(r, \alpha)=\frac{E^2}{\left(1+\epsilon \frac{r^{2}}{L^{2}}\right) rf(r)}.
\label{eq21}
\end{equation}
The variations of function $h(r, \alpha)$ versus its variables is presented in Fig. \ref{fig5-5} by the two panels so that
the right panels show the variations versus orbital radius with $\alpha=0.01$ and the left panels show the variations versus $\alpha$ with $r=1$.
Note that the allowed range for orbital radius is: ($r>r_{h}$ which $r_{h}=0.45$   for $R^{2}$ gravity  and $r_{h}=0.34$ for $R$ gravity).
As the figure shows, the orbital  frequency is an asymptotically decreasing function, in the two gravity theories. But in a quantitative comparison, one realizes that the orbital frequency in Einstein gravity is less than the modified gravity which physically means stronger gravity in modified theory.
\section{Thermodynamic properties }\label{sec-4}
In this section, we examine some important thermodynamic properties of GBH in the modified  and Einstein's frameworks. We will also discuss the thermodynamic stability.
\subsection{Mass}\label{subsec-1}
\begin{figure}[]
\centering
\includegraphics[width=12cm]{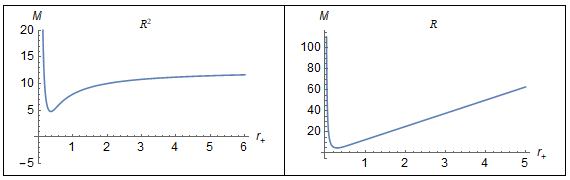}
\caption{The mass variations versus $r_{+}$ in $R^{2}$ and $R$ gravities ((\ref{eq22}) and(\ref{eq23}), respectively) for $\alpha=0.01$.}
\label{fig6-6}
\end{figure}

In the BH thermodynamics, the mass plays the role of internal energy of the BH, therefore, it is important to consider it from the point of view of system energy changes. The conventional way for obtaining the mass function is the equation determining the BH horizon, that is zeroing the metric functions (\ref{eq10}) and (\ref{eq11}) ($f_{R^{2}}(r_{+})=0$ and $f_{R}(r_{+})=0$). By doing this,  the mass functions (in the two theories of gravity) can be obtained as:
\begin{equation}
M_{R^{2}}({r_+})=\frac{1}{\frac{2{{\alpha }^{\frac{1}{2}}}}{{{\pi }^{\frac{3}{2}}}{{r}_{+}}}-\frac{10\alpha}{{{( 4\pi \alpha  )}^{\frac{3}{2}}}}{{\operatorname{e}}^{\frac{-r_{+}^{2}}{4\alpha }}}{{r}_{+}}-\frac{40{{\alpha}^{2}}}{{{( 4\pi \alpha  )}^{\frac{3}{2}}}}\frac{{{\operatorname{e}}^{\frac{-r_{+}^{2}}{4\alpha }}}}{r_{+}}+\frac{1+\frac{12\alpha }{r_{+}^{2}}}{4\pi }Erf[ \frac{{{r}_{+}}}{2\sqrt{\alpha}}]}
\label{eq22}
\end{equation}
and
\begin{equation}
M_{R}({r_+})=\frac{1}{-\frac{1}{4{{\pi }^{\frac{3}{2}}}{{\alpha }^{\frac{1}{2}}}}{{\operatorname{e}}^{\frac{-r_{+}^{2}}{4\alpha }}}+\frac{1}{4\pi {{r}_{+}}}Erf[ \frac{{{r}_{+}}}{\sqrt{4\alpha}}].}
\label{eq23}
\end{equation}
The plots of two  mass functions  versus event horizon are shown in Fig.\ref{fig6-6}, which the left (right)  panel corresponds to modified (Einstein) gravity.
The figure shows in both gravitational theories the mass  is a positive--valued convex function and decreases from large values to a minimum, which means physically that forming the BH is conditional on a minimum mass.
But, after passing the minimal point in the modified theory, unlike the Einstein theory (where the mass tends again towards large values)\footnote{As Schwarzschild BH.}, the mass tends asymptotically to a maximum value.
This means that in the modified theory the mass of the BH cannot exceed a certain limit as its horizon  becomes big and bigger.
This result has more physical plausibility than the Einstein theory which  allows for unlimited increase in  mass.
\subsection{Entropy}\label{subsec-2}
\begin{figure}[H]
\centering
\includegraphics[width=12cm]{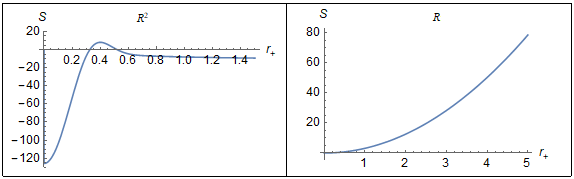}
\caption{The entropy variations versus $r_{+}$ in $R^{2}$ (\ref{eq28}) and $R$ gravities for $\alpha=0.01$.}
\label{fig7-7}
\end{figure}

According to Bekenstein--Hawking formula within the context of modified gravity $F(R)$, the entropy of a BH  is given by \cite{Airy31}:
\begin{equation}
S_{F(R)}=\frac{A}{4}{F}'(R),
\label{eq24}
\end{equation}
where $A=4\pi r_+^{2}$ is the area of event horizon and $F'(R)$ should be computed at event horizon $r_+$ . For  $F(R)=R^2$, formula (\ref{eq24}) becomes
\begin{equation}
S_{R^2}(r_+)=\pi {r_{+}^{2}}R(r_+).
\label{eq25}
\end{equation}
It isn't very hard to show that the Ricci scalar for spherical static symmetry metric (\ref{eq6}) takes the following form:
\begin{equation}
R=R(r)=\frac{r^{2}f''(r)+4rf'(r)+2f(r)-2}{r^{2}},
\label{eq26}
\end{equation}
which by substituting into (\ref{eq25}), one gets
\begin{equation}
S_{R^2}(r_+)=2\pi \Big(f''(r_+)+4rf'(r_+)+2f(r_+)-2\Big).
\label{eq27}
\end{equation}
By replacing the metric function (\ref{eq10}) and its corresponding derivatives into (\ref{eq27}), the entropy function is obtained as
\begin{equation}
S_{R^{2}}(r_{+})=\frac{( 5r_{+}^{5}-48 \alpha r_{+}^{3} +8{\alpha }^{2}{r}_{+} )}{8\sqrt{\pi}{{\alpha }^{5/2}}}M(r_+){{\text{e}}^{-\frac{r_{+}^{2}}{4\alpha }}}-M(r_+)Erf\left[ \frac{{{r}_{+}}}{2\sqrt{\alpha }}\right].
\label{eq28}
\end{equation}
Note that in the latter formula, the mass function $M(r_+)$ must be replaced by formula (\ref{eq22}).
The left panel in Figure \ref{fig7-7}  shows the entropy function (\ref{eq28}) versus horizon radius and right panel shows the corresponding one in Einstein gravity, that is $S_R=\frac{A}{4}=\pi {{r_+}^{2}}$.
The notable point in the figure is  that unlike  the Einstein gravity, where entropy increases quadratically (means, the entropy changes are always positive), in modified gravity, the entropy and its changes can be  positive, zero or negative. Also, note that entropy remains almost (negative) constant beyond a certain horizon  radius.
As we know, the negative entropy is related to the information paradox, suggesting that information may be preserved rather than lost, potentially encoded with negative entropy. Finally, it should be noted that, the BHs which exhibit unusual thermodynamic properties, can experience a decrease in entropy through Hawking radiation during their evaporation process. Therefore, the prediction of negative entropy (or negative entropy changes) by the modified theory can be considered a more complete description than the Einstein theory.\\
\subsection{Temperature}\label{subsec-3}
For the static and spherically symmetric BH spacetime equipped with the metric (\ref{eq6}), the Hawking temperature of the horizon is expressed by the following formula \cite{Airy32}:\\
\begin{equation}
T=\frac{{f}'(r_{+})}{4\pi }.
\label{eq29}
\end{equation}
By calculating ${f}'(r_{+})$ from metric functions (\ref{eq10}) and (\ref{eq11}) and substituting  into (\ref{eq29}), the temperatures (in two gravities) are obtained as follows:
\begin{align}
{{T}_{{R}^{2}}}({{r}_{+}})=\frac{1}{4\pi }\Big(-\frac{5{{\text{e}}^{-\frac{r_{+}^{2}}{4\alpha }}}r_{+}^{2}}{8{({\pi \alpha})^{3/2}}}-\frac{5\sqrt{\alpha }{{\text{e}}^{-\frac{r_{+}^{2}}{4\alpha }}}}{4{{\pi }^{3/2}}}+\frac{2\sqrt{\alpha }}{{{\pi }^{3/2}}r_{+}^{2}} &\nonumber\\ -\frac{5\sqrt{\alpha }{{\text{e}}^{-\frac{r_{+}^{2}}{4\alpha }}}}{{{\pi}^{3/2}}r_{+}^{2}}-\frac{{{\text{e}}^{-\frac{r_{+}^{2}}{4\alpha }}}\left( 1+\frac{12\alpha }{r_{+}^{2}} \right)}{4{{\pi }^{3/2}}\sqrt{\alpha }}+\frac{6\alpha Erf\Big[\frac{{{r}_{+}}}{2\sqrt{\alpha }}\Big]}{\pi r_{+}^{3}}\Big)M(r_+)
 \label{eq30}
\end{align}
and
\begin{equation}
T_{R}({r_+})=\frac{1}{4\pi }\Big(-\frac{M(r_+){{\text{e}}^{-\frac{r_{+}^{2}}{4\alpha }}}{{r}_{+}}}{8{({\pi \alpha})^{3/2}}}+\frac{1}{r_{+}}\Big).
\label{eq31}
\end{equation}
\begin{figure}[H]
  \centering
  \includegraphics[width=10cm]{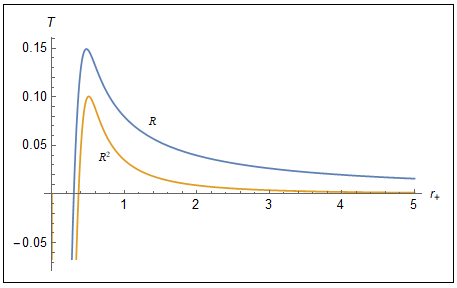}
  \caption{The temperature variations versus $r_{+}$ in $R^{2}$ and $R$ gravities ((\ref{eq30}) and(\ref{eq31}), respectively) in $\alpha=0.01$. The modified gravity model predicts a lower temperature compared to Einsteinian gravity.}
  \label{fig8-8}
\end{figure}
To plot these temperature functions in terms of $r_+$, the mass function $M(r_+)$ must be substituted from equations (\ref{eq22}) and (\ref{eq23}).
The graphs of the temperature functions (\ref{eq30})  and (\ref{eq31}) (versus horizon radius) are shown in Fig. \ref{fig8-8} for $R^2$  and $R$ gravities. The Figure shows that  for $T>0$, the temperature variations in both theories are qualitatively similar. There are  two  points to note:\\
1) The BH temperature has a maximum value meaning that the surface gravity of the Gaussian BH cannot be exceeded from a certain value.\\
2) Quantitatively, for $T>0$, the modified gravity predicts a lower temperature for the BH than the Einstein gravity. Therefore, from the perspective of the lack of radiation from BHs in the present universe, the modified theory may have greater physical acceptability.\\
\subsection{Thermal Stability}\label{section-4}
The HC and GE play an important role in determining the stability of BHs which are usually used to analyze the local and global stability of BHs, respectively. In this subsection, we discuss the stability of the GBH in $R^{2}$ and Einstein  gravities.\newline
\subsubsection{Heat Capacity}\label{section-4}
\begin{figure}[]
\centering
\includegraphics[width=14cm]{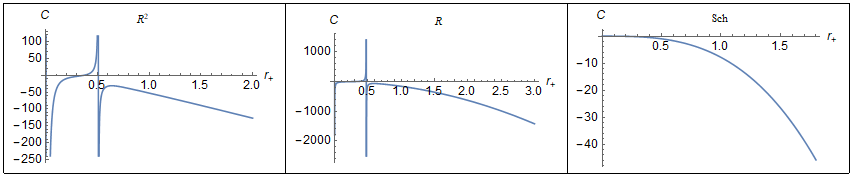}
\caption{The HC variations versus $r_{+}$ in $R^{2}$ (\ref{eq37}), $R$ (\ref{eq38}) gravities and Schwarzschild for $\alpha=0.01$.}
\label{fig9-9}
\end{figure}

In a general BH system, the HC can be defined by:
\begin{equation}
C=\frac{\partial M}{\partial T},
\label{eq32}
\end{equation}
which can also be written as:
\begin{equation}
C=\frac{\partial M}{\partial T}=\left(\frac{\partial M}{\partial r_+}\right)   \left(\frac{\partial r_+}{\partial T}\right) =  \frac{\frac{\partial M}{\partial r_+}}{\frac{\partial T}{\partial r_+}} .
\label{eq33}
\end{equation}
To calculate the HC in modified gravity (for brevity), we put:
\begin{equation}
A=\frac{\partial M}{\partial r_+},\hspace{1cm}  B= \frac{\partial T}{\partial r_+},
\label{eq34}
\end{equation}
which by calculating from equations (\ref{eq22}) and (\ref{eq30}),  we get:
\begin{equation}
A(r_+)=\frac{\partial M}{\partial {{r}_{+}}}=-\frac{\frac{5{{\text{e}}^{-\frac{{{r_{+}}^{2}}}{4\alpha }}}{{r_{+}}^{2}}}{8{({\pi \alpha })^{3/2}}}+\frac{5\sqrt{\alpha }{{\text{e}}^{-\frac{{{r_{+}}^{2}}}{4\alpha }}}}{4{{\pi }^{3/2}}}-\frac{2\sqrt{\alpha }}{{{\pi }^{3/2}}{{r_{+}}^{2}}}+\frac{5\sqrt{\alpha }{{\text{e}}^{-\frac{{{r_{+}}^{2}}}{4\alpha }}}}{{{\pi }^{3/2}}{{r_{+}}^{2}}}+\frac{{{\text{e}}^{-\frac{{{r_{+}}^{2}}}{4\alpha }}}(1+\frac{12\alpha}{{{r_{+}}^{2}}})}{4{{\pi}^{3/2}}\sqrt{\alpha }}-\frac{6\alpha Erf[\frac{r_{+}}{2\sqrt{\alpha}}]}{\pi {{r_{+}}^{3}}}}{{{\Big(-\frac{5{{\text{e}}^{-\frac{{{r_{+}}^{2}}}{4\alpha }}}r_{+}}{4{{\pi}^{3/2}}\sqrt{\alpha}}+\frac{2\sqrt{\alpha}}{{{\pi }^{3/2}}r_{+}}-\frac{5\sqrt{\alpha}{{\text{e}}^{-\frac{{{r_{+}}^{2}}}{4\alpha }}}}{{{\pi}^{3/2}}r_{+}}+\frac{(1+\frac{12\alpha }{{{r_{+}}^{2}}})Erf[\frac{r_{+}}{2\sqrt{\alpha }}]}{4\pi }\Big)}^{2}}}
\label{eq35}
\end{equation}
and
\begin{multline}
B(r_+)=\frac{\partial T}{\partial {{r}_{+}}}=\\
\frac{
   \Big(-64{{\alpha }^{3}}r_{+}+{{\text{e}}^{-\frac{{{r_{+}}^{2}}}{4\alpha }}}r_{+}( 5{{r_{+}}^{6}}-8\alpha {{r_{+}}^{4}} +64{{\alpha }^{2}}{{r_{+}}^{2}}+352{{\alpha}^{3}})-288\sqrt{\pi}{{\alpha }^{7/2}}Erf[\frac{r_{+}}{2\sqrt{\alpha}}]\Big)
}{
    64{({\pi \alpha})^{5/2}}{{r_{+}}^{4}}
}M(r_+).
\label{eq36}
\end{multline}
Given $A$ and $B$ in (\ref{eq35}) and (\ref{eq36}), the HC (\ref{eq33}) in modified gravity reduces to:
\begin{equation}
C_{R^{2}}({r_+})=\frac{A(r_+)}{B(r_+)}.
 \label{eq37}
\end{equation}
By the same calculations from equations (\ref{eq23}) and (\ref{eq31}) and substituting them into equation (\ref{eq33}), the HC in Einstein's gravity becomes:
 \begin{equation}
{{C}_{R}}({{r}_{+}})=-\frac{32{\pi }^{5/2} \alpha{{\text{e}}^{\frac{{{r_{+}}^{2}}}{4\alpha }}}{{r_{+}}^{2}}( -r_{+}( {{r_{+}}^{2}}+2\alpha)+2\sqrt{\pi }{{\alpha }^{3/2}}{{\text{e}}^{\frac{{{r_{+}}^{2}}}{4\alpha }}}Erf[ \frac{r_{+}}{2\sqrt{\alpha }})}{4\sqrt{\alpha }{{r_{+}}^{2}}( -{{r_{+}}^{2}}+\alpha)-\sqrt{\pi }{{\text{e}}^{\frac{{{r_{+}}^{2}}}{4\alpha }}}r_{+}( {{r_{+}}^{4}}-4\alpha{{r_{+}}^{2}} +8{{\alpha }^{2}})Erf[ \frac{r_{+}}{2\sqrt{\alpha }} ]+4\pi {{\alpha }^{5/2}}{{\text{e}}^{\frac{{{r_{+}}^{2}}}{2\alpha }}}Erf{{[ \frac{r_{+}}{2\sqrt{\alpha }}]}^{2}}}.
\label{eq38}
\end{equation}
The left and middle  panels in Fig.\ref{fig9-9} show the graphs of  HC functions (\ref{eq37}) and (\ref{eq38}) (versus horizon radius)  in modify and Einstein gravities, respectively. Also, for the comparison, the HC of   Schwarzschild BH, is shown in  Fig.\ref{fig9-9} (right panel). The figures show, HC graphs are qualitatively similar in the both theories,  except $r_{+}\rightarrow 0$,  for which HC becomes a (negative) finite value in Einstein gravity and  tends to $\pm\infty$  in modified gravity. In both  theories of gravity, the BH can  experience    continuous (type--one: $C=0$ )  and  discontinuous (type--two: $C\rightarrow \infty$ ) phase transition, contrary to Schwarzschild case in which the HC is always non--positive. Finally, for the large and larger  values of the  horizon, the BH becomes more (locally) unstable as Schwarzschild  BH.
\subsubsection{Gibbs Energy}\label{section-3}
\begin{figure}[H]
  \centering
  \includegraphics[width=12cm]{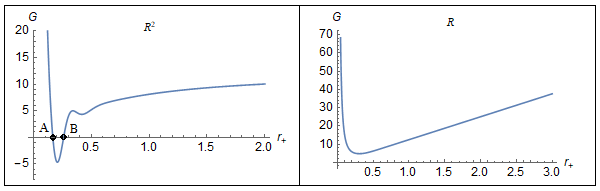}
  \caption{The GE variations versus $r_{+}$ in $R^{2}$ and $R$ gravities ((\ref{eq40}) and(\ref{eq41}), respectively) for $\alpha=0.01$.}
  \label{fig10-10}
\end{figure}

Finally, we use the Gibbs free energy, a useful state function to examine the global stability  and phase transition of  the BH.
In short, positive (negative) GE indicates  global stability (unstability). The transition from negative to positive is accomplished through the equation
$G(r+) = 0$ whose real roots are points of Hawking–-Page phase. In this transition, the BH goes from a stable configuration to an unstable (radiative) one, or vise versa.\\
The GE function introduced by \cite{Airy33}, \cite{Airy34}:
\begin{equation}
G(r_{+})=M-T S,
\label{eq39}
\end{equation}
where  $S, M$ and $T$ are the entropy, mass and temperature, respectively.\\
By substituting $M$, $S$ and $T$ from equations (\ref{eq22}), (\ref{eq28}) and (\ref{eq30}) into (\ref{eq39}), the GB function (in modified gravity) is obtained as:
\begin{equation}
\resizebox{0.9\textwidth}{!}{%
$G_{{R^2}}({r_+}) = \frac{r_{+} \left(32{({\pi \alpha})^{3/2}}{\text{e}}^{\frac{{r_{+}}^{2}}{4\alpha}}r_{+} + \frac{( 5{{r_{+}}^{5}}-48\alpha{{r_{+}}^{3}}+8{{\alpha }^{2}}r_{+}-8\sqrt{\pi }{{\alpha }^{5/2}}{{\text{e}}^{\frac{{r_{+}}^{2}}{4\alpha}}}\operatorname{Erf}[ \frac{r_{+}}{2\sqrt{\alpha }}])( 5{{r_{+}}^{5}}+12\alpha{{r_{+}}^{3}}+64{{\alpha }^{2}}r_{+}-16{{\alpha}^{2}}{{\text{e}}^{\frac{{r_{+}}^{2}}{4\alpha}}}(r_{+}+3\sqrt{\pi \alpha}\operatorname{Erf}[ \frac{r_{+}}{2\sqrt{\alpha}}]))}{-10{{\alpha }^{2}}r_{+}( {{r_{+}}^{2}}+4\alpha)+2{{\alpha}^{5/2}}{{\text{e}}^{\frac{{r_{+}}^{2}}{4\alpha }}}(8\sqrt{\alpha}r_{+}+\sqrt{\pi}({{r_{+}}^{2}}+12\alpha)\operatorname{Erf}[ \frac{r_{+}}{2\sqrt{\alpha}}])}\right)}{8\alpha\left( -5r_{+}( {{r_{+}}^{2}}+4\alpha)+\sqrt{\alpha}{{\text{e}}^{\frac{{r_{+}}^{2}}{4\alpha}}}( 8\sqrt{\alpha }r_{+}+\sqrt{\pi }({{r_{+}}^{2}}+12\alpha)\operatorname{Erf}[ \frac{r_{+}}{2\sqrt{\alpha}}])\right)}$}.
\label{eq40}
\end{equation}
In a similar way, by substituting $S=\pi r_{+}^2$ and $M$, $T$  from equations (\ref{eq23}) and (\ref{eq31}) into (\ref{eq39}), the GB function (in Einstein gravity) is obtained as:
\begin{align}
G_{R}({r_+})=\frac{r_{+}( -r_{+}( {{r_{+}}^{2}}+2\alpha)-2\sqrt{\pi }{{\alpha }^{3/2}}{{\text{e}}^{\frac{{{r_{+}}^{2}}}{4\alpha}}}( -1+16\pi +Erf[\frac{r_{+}}{2\sqrt{\alpha}}]))}{8\alpha( r_{+}-{{\text{e}}^{\frac{{{r_{+}}^{2}}}{4\alpha}}}\sqrt{\pi \alpha}Erf[ \frac{r_{+}}{2\sqrt{\alpha}}])}.
 \label{eq41}
\end{align}
As mentioned above, the real roots of  the equation $G(r+) = 0$ are the Hawking--Page  transition points, however, given the form of the  equations (\ref{eq40}) and (\ref{eq41}),  equation $G(r+) = 0$ cannot be solved analytically. Therefore, to reveal the transition points (and stability regions), the graphs of  the Gibbs functions (\ref{eq40})  and (\ref{eq41}) (versus $r_+$) are plotted in  Fig.\ref{fig10-10}.
As the figure shows, in the Einstein gravity (Fig.\ref{fig10-10}-- right panel), the BH is always globally stable, means that it will never experience the radiation phase which seems a bit strange considering the Hawking radiation theory.
In modified gravity (Fig.\ref{fig10-10} -- left panel), there are two Hawking--Page transition points (approximately, in $r_A=0.18$ and $r_B=0.28$). Thus, for $r_A<r_+<r_B$, the physical BH is in the unstable configuration (radiative phase), and it is globally stable for all values except  the last  inequality.
Compared to the Einstein's gravity, this result seems more reasonable.
Another difference in Fig.\ref{fig10-10} that could physically tip the scales in favor of the modified theory is the limiting case in which the energy in the modified theory (unlike the Einstein's theory) tends to a maximum (finite) limit.
\section{Conclusions}\label{sec-5}
In this manuscript,  the geodesic structure and thermodynamic properties of GBH (in modified $R^2$ and Einstein's gravities) are investigated and compared. A summary of important results can be stated as follows:\\
Part One -- Geodesics:\\
1- For the same conditions, the graphs corresponding to the  geodesic paths  are quantitatively identical in the two gravity theories, but, there is a qualitative difference indicating  that the  gravity is stronger in $R^{2}$ gravity.\\
2- The circular paths of  moving particle out side the BH are unstable orbits.  The orbital frequency of  these  orbits  versus the orbital radius are monotonic decreasing functions (qualitatively identical in the two gravity theories).\\
Part Two -- Thermodynamics: Below, we will separately explain the most important results related to thermodynamic functions.\\
1-- Mass:\\
In the both theory of gravity, the BH mass can not be less than a threshold, meaning that  the formation of  the BH is conditional on a minimum mass. But, in  modified gravity (unlike the Einstein theory in which  the mass can grow unlimited), the mass tends asymptotically to a maximum value. This means that in the modified theory the mass of the BH cannot exceed a certain limit as the horizon becomes big and bigger, indicating more physical plausibility.\\
2-- Entropy:\\
In modified gravity,  the entropy  and its  changes can be  positive, zero or negative (unlike  the Einstein gravity in which they  are always positive).
Remind that the negative entropy (changes)  can experience a decrease in entropy through the Hawking radiation during their evaporation process.  Therefore, the prediction of negative entropy (changes) can be considered a more complete description than the Einstein theory.\\
3-- Temperature:\\
For $T>0$, the two theories predict similar temperature changes, qualitatively. But there is an important quantitative difference that physically accounts for the absence of black hole radiation in the present universe. That is, we know that the reason black holes do not radiate in the present universe is because of their low temperature, so the preference is with the model that predicts (relatively) lower temperatures and is more consistent with the physical world.\\
4-- Heat capacity:\\
Except for $r_{+}\rightarrow 0$, the two theories predict similar  changes, qualitatively.  The BH can go from (locally)  stable configuration to  the unstable one through type--one (continuous) or type--two (discontinuous) phase transitions. Also, after passing a certain horizon radius, the HC  behavior becomes Schwarzschild--like.\\
6- Gibbs Energy:\\
In the Einstein gravity the GE is always positive, indicating viable (globally) stability. In modified gravity, the BH can experience both globally  stable and unstable phase by crossing  the Hawking--Page  transition points. Compared to the Einstein's gravity, this result seems more reasonable.

\end{document}